\documentstyle[aps,preprint,eqsecnum]{revtex}
\begin{document}
\title{Anomalous dimensions and phase transitions in 
superconductors}
\author{Flavio S. Nogueira\thanks{address after september 1st 2000: 
Institut f\"ur Theoretische Physik, Freie Universit\"at Berlin, Arnimallee 
14, D-14195 Berlin}}
\address{Centre de Physique Th\'eorique,  
Ecole Polytechnique, 
F-91128 Palaiseau Cedex, FRANCE}

\date{Received \today}

\maketitle

\begin{abstract}
The anomalous scaling in the Ginzburg-Landau model for the superconducting 
phase transition is studied. It is argued that the negative sign of the 
$\eta$ exponent is a consequence of a special singular behavior in 
momentum space. The negative sign of $\eta$ comes from the divergence 
of the critical correlation function at finite distances. This behavior 
implies the existence of a Lifshitz point in the phase 
diagram. The anomalous scaling of the vector potential is also discussed. 
It is shown that the anomalous dimension of the vector 
potential $\eta_A=4-d$ has important consequences for the critical dynamics 
in superconductors. The frequency-dependent conductivity is shown to 
obey the scaling $\sigma(\omega)\sim\xi^{z-2}$. The prediction $z\approx 3.7$ 
is obtained from existing Monte Carlo data.              
\end{abstract}
\draft

\section{Introduction}

The superconducting phase transition has received considerable attention in 
recent years. All this interest is due in part to the experimentally larger 
critical region in the high-$T_{c}$ materials \cite{Lobb,Salamon,Kamal}. 
This larger critical region, however, does not correspond to the 
inverted 3D XY ($IXY_3$ for short) universality class \cite{Dasgupta}. 
Instead, the observed 
critical behavior belongs to the ordinary 3D XY 
($XY_3$ for short) universality class, meaning 
that the phase transition is governed by the neutral non-trivial 
Wilson-Fisher fixed point. Concerning the charged transition
(that is, the $IXY_3$ behavior), 
there is some progress from the theoretical side. Unfortunatly, the 
corresponding critical region remains experimentally out of reach.

Concerning the $IXY_3$ regime, 
interesting precise numerical results on the anomalous 
scaling dimensions have been obtained recently by Sudb{\o} and 
collaborators \cite{Sudboe,Nguyen,Hove} using a lattice version 
of the Ginzburg-Landau (GL) model. Their results give  
a strong support to the duality scenario 
\cite{Dasgupta,Kleinert,Kiometzis,Kiometzis2}  
which underlies the $IXY_3$ behavior. The aim ot 
this paper is to provide an analysis of the anomalous scaling dimensions 
from the point of view of field theory. 
An important issue to be understood is the sign of the 
order parameter field anomalous dimension, $\eta$. As argued in Ref. 
\cite{deCalan2} a negative $\eta$, though fulfilling the 
inequality $\eta>2-d$, would spoil some important properties that must be 
verified in any legitimate continuum (scaling) limit. 
A fundamental property, the positivity of the spectral weight of the 
K\"allen-Lehmann (KL) spectral representation of the 2-point correlation 
function, is violated if $\eta<0$.
Kiometzis and Schakel \cite{Kiometzis1} pointed out 
also that unitarity is violated if $\eta<0$. In fact, violation of 
the unitarity is an immediate consequence of the violation of the 
positivity of the spectral weight. We should note, however, that most 
renormalization group (RG) calculations give in general $\eta$ 
in the range $-1<\eta<0$ \cite{HLM,Hikami,Radz,Folk,Berg,Herbut1,deCalan1} in 
$d=3$ ($\epsilon=1$ in the context of the $\epsilon$-expansion). 
The only exception is the case where a mass (Proca) term is added 
explicitly for the gauge field, where the inequality $\eta\geq 0$ is 
satisfied \cite{Nogueira}. This last situation corresponds just to the 
case of the continuum dual model where the gauge symmetry is global 
\cite{Kiometzis,Herbut2,Tesanovic,deCalan2}. Since in the RG calculations 
$\eta$ is only slightly negative we may wonder if such a negativeness is 
not just an artifact of the approximations used. The situation is, however, 
much more subtle. The recent numerical simulations in the lattice of 
Nguyen and Sudb{\o} \cite{Nguyen} gives $\eta=-0.18$ \cite{Note1}. The 
results in Ref. \cite{Nguyen} are non-perturbative, in contrast to most 
RG calculations. In the RG context we can cite the work of Bergerhoff 
{\it et al.} \cite{Berg} and the 
$1/N$ expansion \cite{HLM}, both non-perturbative and 
giving also $\eta<0$. 

From a thermodynamical point of view, the anomalous dimension  
$\eta_A$ of the vector potential has more far reaching 
consequences. Indeed, it plays an important role 
in a critical regime where the magnetic fluctuations are not negligible, 
such as in the $IXY_3$ regime. Gauge invariance allows an exact 
determination of $\eta_A$ in $2<d<4$ dimensions. Indeed, its 
value is given simply by $\eta_A=4-d$. One important consequence of 
this result is the scaling $\lambda\sim\xi$ 
\cite{Herbut1,Herbut2,Olsson,deCalan2} where $\lambda$ is the 
penetration depth and $\xi$ is the correlation length.  

In this paper we will discuss some 
interesting new aspects of the superconducting 
transition. We will focus the issue of the anomalous dimensions of fields, 
for both the scalar field and the gauge field.
Our analysis should be 
applicable to superconductors in the type II regime, where we expect a 
second-order (charged) phase transition 
\cite{Dasgupta,Kleinert}.
In section II the negativeness of $\eta$ will be shown 
to be a consequence of the existence of two singularities 
in the scalar 2-point bare correlation function at the critical point (CP). 
One singularity happens at $p=0$ while  
the other one happens at a nonzero momentum $p=p'$. This second singularity 
is related to the existence of a first-order phase transition regime. 
This singularity at nonzero momentum is at the origin of the negative sign 
of the $\eta$ exponent. Indeed, $\eta$ is negative because the order 
parameter wave function renormalization $Z_{\phi}$ is greater than one and 
this happens only if the corresponding {\it critical} 2-point 
correlation function has a pole at nonzero momentum. We argue that this 
behavior implies the existence of a Lifshitz point induced by gauge field 
fluctuations. 

Another point of view is to study the small fluctuations around the 
Halperin-Lubensky-Ma (HLM) mean field theory \cite{HLM}. This is done in 
section III, where the Gaussian fluctuations are calculated in order 
to study the positivity properties of the propagators. It turns out 
that the propagators are positive definite and no pole at $p\neq 0$ is 
found at the CP. Indeed, in order to find out such a pole it is 
necessary to compute the non-Gaussian fluctuations. The analysis of the 
section III shows that the functional integral has a well defined 
Gaussian measure.

In section IV we discuss the physical consequences of the anomalous 
scaling dimension of the magnetic vector potential. After reviewing 
some known properties like the scaling $\lambda\sim\xi$ 
\cite{Herbut1,Herbut2,Olsson,deCalan2}, we analyse the consequences of 
the magnetic fluctuations for the frequency dependent conductivity, 
$\sigma(\omega)$. As argued by Fisher {\it et al.} \cite{Fisher}, in the 
$XY_3$ regime scales as $\sigma(\omega)\sim|t|^{\nu(d-2-z)}$ 
(for the sake of generality we wrote the scaling relation in dimension 
$2<d<4$, that is, a $XY_d$ regime). However, if the magnetic 
fluctuations are included the anomalous dimension of the vector potential 
is no longer equal to zero. This implies the dimension independent 
scaling $\sigma(\omega)\sim|t|^{\nu(2-z)}$. We point out that the scaling 
$\nu'=\nu$ implied by $\lambda\sim\xi$ ($\nu'$ is the penetration depth 
exponent) is also dimension independent, in contrast with the dimension 
dependent result of the $XY_d$ regime, $\nu'=\nu(d-2)/2$. In section V 
we infer from the Monte Carlo data of Lidmar {\it et al.} that 
$z\approx 3.7$ in the $IXY_3$ regime, which is a translation of one 
unity of the result obtained by these authors ($z\approx 2.7$). 
This difference is due to the fact that the scaling 
$\sigma(\omega)\sim|t|^{\nu(d-2-z)}$ was assumed in their Monte Carlo 
simulation of the $IXY_3$ regime. 
Finally, we discuss the relevance of these ideas 
to the Bose-glass transition \cite{Nelson,Lidmar1} in the direction 
perpendicular to the columnar defects, where a transverse Meissner effect 
happens \cite{Lidmar1}.

\section{Phase transitions and the order parameter anomalous dimension}   

In order to fix the ideas, 
let us consider first the case of a scalar $O(2)$ invariant field theory 
with bare Lagrangian

\begin{equation}
\label{O(2)}
L=|\partial_{\mu}\phi|^2+m^2|\phi|^2
+\frac{u}{2}|\phi|^4.
\end{equation} 
Such a theory has a non-trivial 
infrared stable fixed point at $d=3$. The 2-point bare 
truncated correlation function is diagonal in the color indices 
and is defined by 

\begin{equation}
\label{2pf}
W^{(2)}(x,y)=Z^{(2)}(x,y)-\langle\phi(x)\rangle
\langle\phi^{\dag}(y)\rangle, 
\end{equation}
where

\begin{equation}
\label{Z2}
Z^{(2)}(x,y)=\langle\phi(x)\phi^{\dag}(y)\rangle.
\end{equation}  
The 2-point function 
$Z^{(2)}$ has the Fourier representation

\begin{equation}
Z^{(2)}(x,y)=\int\frac{d^d p}{(2\pi)^d}e^{ip\cdot(x-y)}
\tilde{Z}^{(2)}(p),
\end{equation} 
which satisfies the KL spectral representation 
\cite{GJ}: 

\begin{equation}
\label{KL}
\tilde{Z}^{(2)}(p)=c\delta^d (p)+\int_{0}^{\infty}
d\mu\frac{\rho(\mu)}{p^2+\mu^2}, 
\end{equation}
where $\rho$ is the spectral 
density satisfying 

\begin{equation}
\label{sumrule}
\int_{0}^{\infty}d\mu\rho(\mu)=1. 
\end{equation}
From  Eq. (\ref{KL}) we obtain 

\begin{equation}
\label{KL1}
Z^{(2)}(x,y)=c+\int_{0}^{\infty}d\mu\rho(\mu)\frac{e^{-\mu|x-y|}}{
4\pi|x-y|}.
\end{equation}
Let us put $y=0$ for convenience. Then, when the symmetry is broken, 
$W^{(2)}(x,0)\to 0$ as $|x|\to\infty$. Therefore, from Eqs. (\ref{2pf}) and 
(\ref{KL1}) we obtain that $c=|\langle\phi(0)\rangle|^2$ which is 
different from zero if $T<T_c$, vanishing otherwise. 

Using Eq. (\ref{sumrule}) it  
follows easily that the Fourier transform of the bare truncated 
2-point correlation function satisfies the 
infrared bound \cite{Spencer},  
$\tilde{W}^{(2)}(p)\leq 1/p^2$. Moreover, 
Griffiths correlation inequality \cite{GJ} implies 
$\tilde{W}^{(2)}(p)\geq 0$. Therefore, 

\begin{equation}
\label{ineq1}
0\leq\tilde{W}^{(2)}(p)\leq\frac{1}{p^2}.
\end{equation}
The inequality (\ref{ineq1}) has an important consequence for the infrared 
behavior. At the CP, the bare correlation function behaves as 
$\tilde{W}^{(2)}(p)\sim 1/p^{2-\eta}$ as $p\to 0$ and Eq. (\ref{ineq1}) 
implies therefore that $\eta\geq 0$. Note that in the above argument 
no reference is made to the global character of the symmetry group. Thus, 
we may think that the same rule should apply to the GL model where the 
gauge symmetry is local. We will see that this is not the case.     

The bare Lagrangian of the GL model is

\begin{equation}
\label{L}
L=\frac{1}{4}F^2+(D_{\mu}\phi)^{\dag}
(D_{\mu}\phi)
+m^2|\phi|^2
+\frac{u}{2}|\phi|^4,
\end{equation}
where $F^{2}$ is a short for 
$F^{\mu\nu}F^{\mu\nu}$, $F^{\mu\nu}=\partial_{\mu}A_{\nu}
-\partial_{\nu}A_{\mu}$,  
and $D_{\mu}=\partial_{\mu}+ieA_{\mu}$.  
At 1-loop, we obtain for $d=3$, $T\geq T_{c}$ and in the Coulomb gauge 
$\partial_{\mu}A_{\mu}=0$,   

\begin{equation}
\label{2pfGL}
\tilde{W}^{(2)}(p)=\frac{1}{p^2+m^2+\Sigma(p)},
\end{equation}
with the self-energy

\begin{equation}
\label{SE}
\Sigma(p)=-\frac{m}{2\pi}(u+e^2)-
\frac{e^2}{4\pi|p|}(p^2-m^2)
\left[\frac{\pi}{2}+
\arctan\left(\frac{p^2-m^2}{2m|p|}\right)
\right].
\end{equation}
In writing the above equations we have absorbed in the bare mass
a contribution with a linear dependence on the ultraviolet cutoff 
$\Lambda$. Thus, $m^2\propto t$, where $t=(T-T_c)/T_c$ is the 
reduced temperature. The correlation function 
$\tilde{W}^{(2)}(p)$ at $p=0$ gives the susceptibility $\chi$. The 
divergence of the susceptibility at $T=T_c$ signals a phase transition. 
In terms of the correlation length $\xi=m_r^{-1}$, 
where $m_r$ is the renormalized mass, the susceptibility is 
written as $\chi=Z_{\phi}\xi^2$, where $Z_{\phi}$ is the wave-function 
renormalization. Here is the crucial point. For the $O(2)$ model, the 
2-point correlation function diverges at $T_{c}$ {\it only} for $p=0$. 
The same is not true for the GL model. In fact, the above 1-loop 
calculation shows that for $|p|=p'=e^2/4$ the 2-point correlation 
function also diverges at $T_c$. Thus, we can define a second 
susceptibility $\chi'=\tilde{W}^{(2)}(p')$. The existence of a 
second pole in $\tilde{W}^{(2)}$ implies that $Z_{\phi}>1$. 
Thus, the infrared bound Eq. (\ref{ineq1}) does not hold. If moreover 
we assume that the phase transition at $p=0$ is of second-order, we obtain that 
$\eta<0$. The same result holds at 2-loops and also in the $1/N$ expansion. 
A negative value of 
$\eta$ is also found by means of non-perturbative RG \cite{Berg} and 
in a recent Monte Carlo simulation \cite{Nguyen}. This strange behavior 
needs an explanation and an interpretation. 
Note that not only the right hand side 
of (\ref{ineq1}) is violated but also its left hand side. 
The striking feature of this behavior is that 
$|\tilde{\phi}(p)|^2\geq 0$ for all $p$ but 
$\tilde{W}^{(2)}(p)=\langle|\tilde{\phi}(p)|^2\rangle<0$ if 
$0<|p|<p'$. The average of the everywhere positive operator 
$|\tilde{\phi}(p)|^2$ is not positive everywhere! 
Thus, it seems that the corresponding effective 
Gaussian measure is not positive definite and, 
as a consequence, the functional integral is not well-defined. Of 
course, the KL representation cannot hold with a positive measure.   

Let us explain the meaning of the susceptibility $\chi'$. The fact that 
$\tilde{W}^{(2)}(p')$ diverges at $T_c$ means that a phase transition 
happens at finite distances. 
This is a typical feature of a first-order phase transition.  
The first- and second-order phase transition can be described at a same 
$T_c$ but at different momentum scales, $p=0$ for the second-order 
phase transition and $|p|=p'$ for the first-order one. This shed new 
light in the RG fixed dimension approach at the 
CP of Refs. \cite{Herbut1,deCalan1}, 
where two momentum scales are considered, defining in this way two 
characteristic lengths (note that for $T<T_c$ there are two lengths 
in the problem, namely, the correlation length $\xi$ and the penetration 
depth $\lambda$). For $T=T_c$, the fixed point structure is such that 
both phase transition regimes are contained in the RG flow diagram 
determined by dimensionless couplings $\hat{u}(\mu)=u_{r}(\mu)/\mu$ 
and $\hat{e}^2(\mu)=e_r^2(\mu)/\mu$, 
with $u_r$ and $e_r$ being the renormalized counterparts of $u$ and $e$.  
The regions of first- and second-order phase transition are separated by a 
line connecting the Gaussian fixed point and the so called tricritical 
fixed point \cite{Note}. 
This fixed point is infrared stable along the tricritical 
line and unstable in the the direction of $\hat{u}$. For momentum scales  
such that the couplings are at the left of the tricritical line, the 
phase transition is of first-order. Concerning the sign of $\eta$, it 
must be observed the following crossovers. The first one corresponds to 
zero charge, $\hat{e}^2=0$. In this case the flow is towards the 
$XY_3$ fixed point and $\eta\geq 0$ ($\eta=0$ at 1-loop). This situation 
is  consistent with 
the infrared bound (\ref{ineq1}). The other crossover corresponds 
to the case where the couplings are over the tricritical line. In this 
situation the flow is towards the tricritical point. Both crossovers 
give a critical behavior consistent with a second-order like phase transition. 
It must be stressed, however, that the {\it true} second-order phase 
transition is governed by the infrared stable fixed point. The  
described crossovers are infrared stable only along the crossover lines, 
the tricritical line and the line $\hat{e}^2=0$. The critical regime 
associated to the tricritical line leads to $\eta<0$, in contrast to the 
$XY_3$ crossover. 

The singularity at $|p|=p'$ can be interpreted in terms of  
the effective action. We will write the effective 
action in momentum space rather than in real space. Thus, if $\varphi$ and 
$a_{\mu}$ are the respective 
Legendre transformed fields of $\phi$ and $A_{\mu}$, 
we have $\Gamma=\int d^3p/(2\pi)^3\tilde{\Gamma}(p)$, with

\begin{eqnarray}
\label{effaction}
\tilde{\Gamma}(p)&=&\frac{1}{2}\tilde{\Gamma}^{(2)}(p)\tilde{\varphi}_i(p)
\tilde{\varphi}_i(-p)+\frac{1}{2}\tilde{\Gamma}_{\mu\nu}^{(2)}(p)
\tilde{a}_{\mu}(p)\tilde{a}_{\nu}(-p)\nonumber\\
&+&\frac{1}{4}
\int\frac{d^3 q}{(2\pi)^3}\int\frac{d^3 k}{(2\pi)^3}\tilde{\Gamma}^{(4)}(
p,q,p-k,q+k)(\delta_{ij}\delta_{kl}+\delta_{ik}\delta_{jl}+
\delta_{il}\delta_{jk})\tilde{\varphi}_i(p)\tilde{\varphi}_j(q)
\tilde{\varphi}_k(p-k)\tilde{\varphi}_l(q+k)\nonumber\\
&+&\int\frac{d^3 k}{(2\pi)^3}\tilde{\Lambda}_{\mu}(p-k,p,k)
\tilde{a}_{\mu}(p-k)\tilde{\varphi}_1(p)\tilde{\varphi}_2(k)
\nonumber\\
&+&\frac{1}{2}\int\frac{d^3 q}{(2\pi)^3}\int\frac{d^3 k}{(2\pi)^3}
\tilde{\Omega}(p,q,p-k,q+k)\tilde{\varphi}_i(p)\tilde{\varphi}_i(q)
\tilde{a}_{\mu}(p-k)\tilde{a}_{\mu}(q+k)+(h.o.t.),
\end{eqnarray}
where summation over repeated latin and greek indices is implied and 
we have written $\varphi=(\varphi_1+i\varphi_2)/\sqrt{2}$. Since 
$\tilde{\Gamma}^{(2)}(p)=1/\tilde{W}^{(2)}(p)$, we have that 
at the CP the first term of the RHS of Eq. (\ref{effaction}) vanishes 
when $|p|=p'$ and is negative when $0<|p|<p'$. Of course, when 
$\tilde{\Gamma}^{(2)}(p)$ is negative we must have a positive 
$\tilde{\Gamma}^{(4)}$ to ensure the stability of the effective action. 
In this paper we will not enter into the details of the stability 
conditions with respect to the 4-point function. 

The physical 
picture that emerges from the behavior of the 2-point function is that 
of a tricritical Lifshitz point \cite{Hornreich,Selke,Folk1,Mergulhao}. 
In fact, in 
scalar models for the Lifshitz points the 2-point function vanishes at 
the CP for a nonzero momentum value. In pure scalar models this can happen 
only if higher derivative Gaussian terms are present already at the 
tree level \cite{Hornreich}. Remarkably, in the GL model 
{\it the Lifshitz point 
is induced by the gauge field fluctuations} \cite{Note3}. The existence 
of a tricritical point in the GL model was established by Kleinert 
\cite{Kleinert} using a disorder field theory obtained from duality 
arguments. In the disorder field theory scenario, an effective local 
scalar Lagrangian with disorder parameter $\psi$ is constructed. It has 
been shown that the effective quartic coupling in this model changes 
sign at some point in the coupling space of the original model. This 
characterizes an ordinary tricritical behavior in the disorder field 
theory. In the original GL model this tricritical point is of a 
Lifshitz type and that is the physical interpretation of the negative 
sign of $\eta$. Note that $\eta_{dual}$ is positive in the disorder 
field theory \cite{Kiometzis,Herbut2,deCalan2}. 
In scalar theories of the Lifshitz point the sign of $\eta$ is 
negative in dimensions $d_c-1$ where $d_c$ is the critical dimension 
of the model. For instance, a fixed dimension calculation in a $1/N$ 
expansion gives for the isotropic Lifshitz point in $d=7$ ($d_c=8$ in 
this case) $\eta_{l4}\approx -0.08/N$ \cite{Hornreich1}. 

A Lifshitz point behavior implies the existence of a modulated regime for 
the order parameter. This modulated regime should correspond to the 
type II regime and is analogous to the helical phase in scalar models 
of the Lifshitz point. The type I regime is analogous to the ferromagnetic 
or uniform order parameter phase in these models, the normal regime 
being the analog of the paramagnetic phase. The phase diagram should be  
therefore quite similar to the phase diagram of the 
R-S model \cite{Redner}. The phase diagram of the R-S model is drawn in the 
$T-X$ plane where $X=S/R$ is the ratio between the couplings $S$ and $R$. 
In this phase diagram the line separating the helical phase from the 
ferromagnetic phase is a first order line. In the case of superconductors 
we should draw the phase diagram in a $T-\kappa^2$ plane, where 
$\kappa^2=u/2e^2$ is the square of the Ginzburg parameter.  
The phases paramagnetic, ferromagnetic and helical of the R-S 
model are replaced respectively by normal, type I and type II. Experimentally, 
the modulated nature of the order parameter in the type II regime is 
seen upon applying an external magnetic field and corresponds to the 
Abrikosov vortex lattice \cite{Abrikosov}.

\section{Wave function renormalization from fluctuations around the 
Halperin-Lubensky-Ma mean field theory}

The HLM mean-field theory \cite{HLM} neglects the order parameter 
fluctuations while including the gauge field fluctuations. For an 
uniform order parameter, the gauge field is integrated out exactly and 
a term proportional to $|\phi|^3$ with negative sign is generated in the 
free energy. The corresponding phase transition is 
found to be wekly first-order. RG 
calculations using the $\epsilon$-expansion confirms this scenario since 
no charged fixed point arises. A stable flow towards the infrared happens 
only at zero charge and the $XY_3$ regime follows by taking $\epsilon=1$. 
The $XY$ fixed point is unstable for arbitrarily small charge. This 
behavior remains even at 2-loop order \cite{Kol}. Charged fixed points 
are obtained only by considering an order parameter with $N/2$ complex 
components and in the limit of $N$ sufficiently large. Indeed, at 
1-loop order charged fixed points are obtained if $N>365.9$. Interestingly, 
the critical value of $N$ decreases considerably already at 2-loops 
\cite{Kol} and charged fixed points are found for $N>36$. More recently, 
by using Pad\'e-Borel resummation of the $\epsilon$-expansion, 
Folk and Holovatch \cite{Folk} succeeded in obtaining charged fixed 
points for the physical value $N=2$.        

In this section we will evaluate the Gaussian fluctuations around 
the HLM mean-field theory. These fluctuations will not suffice for 
changing the order of the transition, and so it will remains first-order. 
Our interest here is the positivity properties of the 2-point correlation 
function in this fluctuation-corrected Gaussian approximation. 
This amounts in calculating the propagators associated to the HLM mean-field 
solution. Once this is done, the Gaussian 
measure necessary to compute the non-Gaussian fluctuations is determined. 
If this measure is not positive definite, then the functional integral is 
not well defined and all the theory is inconsistent. We will see that this 
is not the case.     

Let us write $\phi=(\phi_1+i\phi_2)/\sqrt{2}$. 
By integrating out exactly the gauge field we obtain,       

\begin{equation}
\label{partition}
Z=\lim_{a\to 0}\int D\phi_1D\phi_2\exp(-S_{eff}),
\end{equation}
where the effective action 

\begin{eqnarray}
\label{Seff}
S_{eff}&=&\frac{1}{2}Tr\ln[\hat{M}_{\mu\nu}(x-y;a)]\nonumber\\
&-&\frac{e^2}{2}\int d^3 x\int d^3 y[\phi_1(x)\partial_{\mu}^{x}
\phi_2(x)-\phi_2(x)\partial_{\mu}^{x}\phi_1(x)]\hat{D}_{\mu\nu}(x-y;a)
[\phi_1(y)\partial_{\nu}^{y}
\phi_2(y)-\phi_2(y)\partial_{\nu}^{y}\phi_1(y)]\nonumber\\
&+&\int d^3 x\left[\frac{1}{2}\phi_1(-\Delta+\delta m^2+m^2)\phi_1
+\frac{1}{2}\phi_2(-\Delta+\delta m^2+m^2)\phi_2
+\frac{u}{8}(\phi_1^2+\phi_2^2)^2\right],
\end{eqnarray} 
where we have introduced a mass counterterm $\delta m^2$, necessary 
to cancel tadpole divergences (see below). The operator 
$\hat{D}$ is the inverse of $\hat{M}$, the latter being given by 

\begin{equation}
\hat{M}_{\mu\nu}(x-y;a)=\delta^3(x-y)\{[-\Delta+
e^2(\phi_1^2+\phi_2^2)]\delta_{\mu\nu}+(1-1/a)\partial_{\mu}\partial_{\nu}\},
\end{equation}
where $a$ is the gauge fixing parameter. In Eq. (\ref{partition}), the 
limit $a\to 0$ is taken in order to inforce the Coulomb gauge condition. 
Now, we consider small fluctuations around $\phi_i=v\delta_{i1}$, 
where $v=const$ is the solution of 

\begin{equation}
\label{sp}
\frac{\delta S_{eff}}{\delta\phi_i}=0.
\end{equation} 
In this case it is legitimate to truncate $S_{eff}$ up to quadratic 
order in the fluctuating fields $\delta\phi_1$ and 
$\delta\phi_2$. The result is 

\begin{eqnarray}
\label{Seff1}
S_{eff}&=&S_{eff}^{HLM}\nonumber\\
&+&\frac{1}{2}\int d^3x\int d^3y\{\delta\phi_1(x)[
(-\Delta+\delta m^2+3\bar{m}^2+m^2+e^2D_{\mu\mu}(0))\delta^3(x-y)\nonumber\\
&-&2e^2M^2D_{\mu\nu}(x-y)
D_{\nu\mu}(y-x)]\delta\phi_1(y)\nonumber\\
&+&\delta\phi_2(x)[(-\Delta+\delta m^2+\bar{m}^2+
+m^2+e^2D_{\mu\mu}(0))\delta^3(x-y)
-M^2\partial_{\mu}^{x}\partial_{\nu}^{y}D_{\mu\nu}(x-y)]\delta\phi_2(y)\},
\end{eqnarray}
where $\bar{m}^2=uv^2/2$, $M^2=e^2v^2$ and  
$S_{eff}^{HLM}$ corresponds to the HLM  
mean-field free energy \cite{HLM}. Also,  

\begin{equation}
D_{\mu\nu}(x-y)=\int\frac{d^3p}{(2\pi)^3}e^{ip\cdot(x-y)}
\tilde{D}_{\mu\nu}(p) 
\end{equation}
is the operator $\hat{D}$ for $\delta\phi_1=\delta\phi_2=0$. 
In the Coulomb gauge, 

\begin{equation}
\tilde{D}_{\mu\nu}(p)=\frac{1}{p^2+M^2}\left(\delta_{\mu\nu}-\frac{
p_{\mu}p_{\nu}}{p^2}\right), 
\end{equation}
which implies

\begin{equation}
\int d^3x\int d^3y\partial_{\mu}^{x}\partial_{\nu}^{y}D_{\mu\nu}(x-y)
\delta\phi_2(x)\delta\phi_2(y)=0.
\end{equation}
Now we see that the counterterm $\delta m^2$ is necessary in order to 
cancel the linear cutoff dependence coming from the tadpole term 
$e^2D_{\mu\mu}(0)$. Therefore, the $\delta\phi_1$ propagator is 

\begin{equation}
G_{11}(p)=\frac{1}{p^2+m^2+3\bar{m}^2+\Sigma_{11}(p)},
\end{equation}
where the self-energy $\Sigma_{11}(p)$ is given by

\begin{equation}
\label{SE11}
\Sigma_{11}(p)=e^2D_{\mu\mu}(0)-2e^2M^2\int\frac{d^3k}{(2\pi)^3}
\tilde{D}_{\mu\nu}(k-p)\tilde{D}_{\nu\mu}(k).
\end{equation}
By evaluating the integrals in Eq. (\ref{SE11}) we obtain 

\begin{eqnarray}
\label{S11}
\Sigma_{11}(p)&=&-\frac{e^2M}{2\pi}-e^2\left[\frac{M}{4\pi}-\frac{p}{8}
-\frac{M^2}{16p}+\frac{p^4+8M^4+4M^2p^2}{8\pi pM^2}
\arctan\left(\frac{p}{2M}\right)\right.\nonumber\\
&-&\left.\frac{(p^2+M^2)^2}{8\pi p M^2}\arctan\left(\frac{p^2-M^2}{2pM}
\right)\right].
\end{eqnarray}
The $\delta\phi_2$ propagator is given simply by 

\begin{equation}
G_{22}(p)=\frac{1}{p^2+m^2+\bar{m}^2-\frac{e^2M}{2\pi}}.
\end{equation}

Note that the above calculation differs from the usual 1-loop result. In an 
ordinary 1-loop calculation we integrate out the quadratic fluctuations 
around the solution $v=(-2m^2/u)^{1/2}$ corresponding to the tree-level 
with ${\bf A}=0$. Above, we integrated out ${\bf A}$ first and then we 
computed the Gaussian fluctuations around a solution $v$ given by 
Eq. (\ref{sp}), which already contain magnetic fluctuations.
Eq. (\ref{sp}) have the following solutions: 

\begin{equation}
v=0,
\end{equation}

\begin{equation}
\label{sols}
|v|=\frac{e^3}{2\pi u}\pm\frac{1}{u}\sqrt{\frac{e^6}{4\pi^2}-2um^2}.
\end{equation}
Thus, in the ordered phase given by Eq. (\ref{sols}) we have 
$m^2+\bar{m}^2-e^2 M/2\pi=0$ and therefore the $\delta\phi_2$ propagator 
$G_22(p)$ is massless. 
Then, the fluctuating field $\delta\phi_2$ is the would-be 
Goldstone boson of the theory.  
When $m^2=e^6/(8\pi^2 u)$ the square root in Eq. (\ref{sols}) 
vanishes. This value of $m^2$ corresponds to a point of non-analyticity 
of $v$ as a function of $m^2$. Indeed, the derivative of $v$ with respect 
to $m^2$ diverges for $m^2=e^6/(8\pi^2 u)$. Thus, if we  
expand the denominator of $G_{11}$ for $p$ small we obtain, 

\begin{equation}
\label{G11expand}
G_{11}(p)=\frac{1}{\left(1+\frac{5}{24\pi}\frac{e^2}{M}\right)p^2+m^2+
3\bar{m}^2-\frac{e^2 M}{\pi}+O(p^4)}, 
\end{equation} 
and we see that the susceptibilities 
$\chi_i=G_{ii}(0)$ ($i=1,2$) diverge together if $m^2=e^6/(8\pi^2 u)$. 
This singular behavior of the susceptibilities {\it is not} associated to 
any phase transition. It is just an artifact of our fluctuation-corrected
Gaussian approximation. The singularity of $\chi_i$ for $m^2=e^6/(8\pi^2 u)$ 
is inherited from the non-analytic behavior of $v$ for this value of $m^2$. 
Once the non-Gaussian fluctuations are taken into account and a full 
renormalization of mass and coupling constants is done, this artifact 
disappears. On the other hand, if $m^2<e^6/(8\pi^2 u)$, $\chi_2$ diverge 
but not $\chi_1$.  
In this fluctuation induced phase transition scenario $v=\langle\phi\rangle$ 
is different from zero at the CP, a typical behavior of a first-order 
transition, as we have already discussed in section II. 
Note that in this calculation the 
correlation functions diverge only at $p=0$. As a consequence, the 
wave function renormalizations $Z_i\leq 1$. 
By putting $v=e^3/(\pi u)$ which corresponds 
to $m^2=0$ in Eq. (\ref{G11expand}), we obtain  

\begin{equation}
\label{Z1}
Z_1=\frac{1}{1+\frac{5}{12}\kappa^2}<1. 
\end{equation} 
Therefore, the corresponding Gaussian measure is positive definite and 
the non-Gaussian fluctuations can be calculated by means of this 
measure.  The non-Gaussian fluctuations will ultimately make $Z_1>1$, 
violating again the KL representation.   

\section{The vector potential anomalous dimension and its consequences for 
the critical dynamics in superconductors}

One important feature of the $IXY_3$ universality class is the scaling 
$\lambda\sim\xi$ \cite{Herbut1,Herbut2,Olsson,deCalan2}, where 
$\lambda$ and $\xi$ are the penetration depth and correlation length, 
respectively. This scaling contrast with the $XY_3$ behavior, where 
$\lambda\sim\xi^{1/2}$ \cite{Lobb,Fisher}. The reason for this different 
behavior comes from the magnetic fluctuations, which in the $XY_3$ 
universality class play no role. In the $XY_3$ 
regime the magnetic vector potential has no anomalous dimension. 
Concerning the scaling of the penetration depth, it was argued in Refs. 
\cite{Herbut1,deCalan2} that the vector potential anomalous dimension 
contributes in such a way that we have 
in general $\lambda\sim\xi^{(\eta_A+d-2)/2}$. 
Thus, when the magnetic fluctuations are negligeable we have $\eta_A=0$ and 
$\lambda\sim\xi^{(d-2)/2}$, implying in this way a penetration depth 
exponent $\nu'=\nu(d-2)/2$ with $\nu\approx 2/3$ when $d=3$. On the other 
hand, if we take into account the magnetic fluctuations, we have that 
$\eta_A=4-d$ and $\lambda\sim\xi$ implying $\nu'=\nu$. The critical 
exponent $\nu$ is the same in both $XY_3$ and $IXY_3$ 
universality classes 
\cite{Kiometzis,Herbut2,deCalan1,deCalan2,Tesanovic,Olsson} and 
we obtain that $\nu'\approx 1/3$ and $\nu'\approx 2/3$ for the 
$XY_3$ and $IXY_3$ regimes, respectively. Note that only the thermodynamic 
exponents coincide in these two $XY$ regimes. As we have already seen, the 
anomalous dimensions are not the same.

The frequency-dependent conductivity $\sigma(\omega)$ 
scales differently in a magnetic fluctuation regime. 
For $T<T_c$ we have that $\sigma(\omega)\sim e^2\rho_s/(-i\omega)$, 
where $\rho_s$ is the superfluid density. Near a charged fixed point we 
have $e^2\sim\xi^{-\eta_A}$ and therefore, 

\begin{equation}
\label{conduc}
\sigma(\omega)\sim\xi^{2-d+z-\eta_{A}},
\end{equation}
where $z$ is the dynamical exponent and we have used the Josephson 
relation $\rho_s\sim\xi^{2-d}$ 
\cite{Josephson,MEFisher,deCalan2}. Again, by neglecting the magnetic  
fluctuations we recover the usual scaling 
\cite{Fisher}. The $XY$ scaling proposed by Fisher {\it et al.} 
\cite{Fisher} was verified 
recently by Wickham and Dorsey \cite{Dorsey}, who calculated $\sigma(\omega)$ 
using the Kubo formula to $O(\epsilon^2)$ in the $\epsilon=4-d$-expansion. 

Since in the magnetic fluctuation regime $\eta_{A}=4-d$, we obtain

\begin{equation}
\label{conduc1}
\sigma(\omega)\sim\xi^{z-2},
\end{equation}
a result independent of the dimension. This independence of the 
dimension in the scaling behavior (\ref{conduc1}) seems to be a special 
feature of the charged fixed point. Note that already in the case of   
the penetration depth we have obtained $\nu'=\nu$ instead of the dimension 
dependent result $\nu'=\nu(d-2)/2$ of the $XY$ regime. The scaling 
given in Eq. (\ref{conduc1}) has been obtained before 
by Mou \cite{Mou} who used a completely different argument. Our 
argument is much more simple and follows from the {\it exact} value of the 
vector potential anomalous dimension. However, the dynamical exponent 
$z$ is not be the same as in the uncharged model, as was claimed 
in Ref. \cite{Mou}. The Monte Carlo simulations 
of Lidmar {\it et al.} \cite{Lidmar} show very 
clearly that this is not the case 
and that the value of $z$ is enhanced by magnetic fluctuations. 
However, Lidmar {\it et al.} fitted their Monte Carlo data to 
$\sigma\sim|t|^{\nu(d-2-z)}$ instead of using the scaling (\ref{conduc1}). 
Since $\eta_A=1$ in $d=3$, we conclude that the numerical result 
of Ref. \cite{Lidmar} should be shifted to obtain $z\approx 3.7$ instead 
of $z\approx 2.7$. This surprisingly high value of $z$ could be, however, a 
matter of controversy. It may be a consequence of the way the authors 
of Ref. \cite{Lidmar} modelled the $I-V$ characteristics of the $IXY_3$ 
regime. For instance, Amp\`ere's law is neglected in their approach. 
 
From the experimental side, the work of Booth {\it et al.} \cite{Booth} 
fit reasonably the value $z=2.7$ but they assume also a scaling with 
$\eta_A=0$. Anyway, in the case of Ref. \cite{Booth} it is more probable 
that the critical region probed does not correspond to a $IXY_3$ 
universality class. In this case $\eta_A=0$ would be a legitimate 
assumption. 

The scaling Eq. (\ref{conduc1}) is also relevant in other situations. For 
example, in the Bose-glass transition the conductivity perpendicular 
to the columnar defects is argued to obey a scaling exactly as in 
Eq. (\ref{conduc1}) \cite{Lidmar1,Nelson}. Although Eq. (\ref{conduc1}) is 
a zero field scaling, it should apply in the nonzero field situation of 
the Bose-glass transition in the direction perpendicular to the 
columnar defects but not in the longitudinal direction. The reason for this 
behavior comes from the fact that in the perpendicular direction a transverse 
Meissner effect happens, implying in this way a zero field like 
situation. The Bose-glass transition is an example which  
shows that at high fields the magnetic 
thermal fluctuations may be experimentally important and observable.

\section{Conclusion}

In this paper we have discussed some new features of the superconducting 
phase transition. The important role of the anomalous dimensions has 
been emphasized. However, the critical behavior disussed in this paper 
is relevant only near a charged fluctuation critical regime. The relevance 
of the ideas discussed here to high-temperature superconductors (HTSCs) 
may be questioned. Usually the HTSCs have very high values of $\kappa$, 
typically in the range $70-100$. For this reason, 
it is generally assumed that magnetic fluctuations do not play an 
important role. This is in fact the case in the {\it extreme} type II 
limit, that is, $\kappa\to\infty$. For extreme type II superconductors, 
the local magnetic induction equals the applied magnetic field and 
the constraint $\nabla\times{\bf A}={\bf H}$ applies \cite{Tesanovic}. 
In the presently accessible critical region, the HTSCs seem to be 
well approximated by an extreme type II limit. In this case the 
$XY_3$ regime dominate at zero or low magnetic fields. 
The $XY_3$ behavior has been probed with considerable confidence in 
YBa$_{2}$Cu$_{3}$O$_{7-\delta}$ (YBCO) crystal samples 
\cite{Salamon,Kamal,Junod1}.  For Bi$_{2}$Sr$_{2}$CaCu$_{2}$O$_{8+\delta}$ 
(BSCCO), however, the situation is less clear due to the experimental 
difficulties involved. Specific heat measurements seem to indicate 
that the universality class is not $XY_{3}$ \cite{Junod2}. The apparent 
failure of the $XY_{3}$ scaling in BSCCO seems also to be corroborated 
by the penetration depth data \cite{Jacobs}. However, inhomegeneities and 
finite size effects can play a significant role in BSCCO and it may 
happen that it obeys also a $XY_{3}$ scaling \cite{Schneider}. 

The $IXY_3$ behavior, on the other hand, seems to be not presently accessible. 
In fact, penetration depth data from YBCO fulfill very well the 
scaling relation $\nu'=\nu/2$ \cite{Kamal}, agreeing with the 
$XY_3$ behavior. Thus, in order to check the theoretical predictions 
concerning the $IXY_3$ regime, we have to compare these mainly to Monte Carlo 
simulations. For instance, the scaling relation $\nu'=\nu$ with 
$\nu\approx 2/3$ was well verified by Olsson and Teitel \cite{Olsson}. The 
value $\eta\approx -0.18$ was obtained by Nguyen and Sudb\o 
\cite{Nguyen}. The dynamical exponent $z$ was studied by Lidmar {\it et al.} 
\cite{Lidmar} both in the $XY_3$ and in the $IXY_3$ regimes. 
However, as discussed in section IV, they assumed 
the same scaling for the frequency dependent conductivity in both regimes. 
This does not invalidate their data, which remain useful and 
lead to the prediction $z\approx 3.7$ instead of $z\approx 2.7$.
While presently there is little 
hope in checking these predictions in zero field experiments, further 
Monte Carlo simulations can be done in order to obtain a definitive 
answer. As far as real experiments are concerned, we have pointed out that 
the scaling given in Eq. (\ref{conduc1}) holds for the condcutivity 
perpendicular to the columnar defects in a Bose-glass transition 
\cite{Nelson,Lidmar1}. Unfortunately, in this nonzero field regime we are 
unable to estimate the value of $z$ with the arguments presented in this 
paper. It is worth to mention, however, that an experimental value 
$z\approx 5.3$ was probed recently by Klein {\it et al.} \cite{Klein} 
for the Bose-Glass transition in the fully isotropic compound 
(K,Ba)BiO$_3$ with columnar defects.  
     
\acknowledgments

The author would like to acknowledge A. Sudb{\o} for sending the 
paper Ref. \cite{Hove} prior to publication. Much of the present 
work originated from discussions with A. Sudb{\o} and Z. 
Te\u{s}anovi\'c and the author is indebted to them. The author  
acknowledges also C. de Calan for numerous discussions.


\begin{references}
\bibitem{Note} The existence of a tricritical point was predicted by 
H. Kleinert; see Ref. \cite{Kleinert}.
\end{references}
\end{document}